\documentclass{PoS}
\usepackage[authoryear,square]{natbib}
\bibpunct{(}{)}{;}{a}{}{,}

\title{Filaments of the radio cosmic web: opportunities and challenges for SKA}
\ShortTitle{Filaments with the SKA}
\author{\speaker{Franco Vazza}$^{1,2}$, 
             Chiara Ferrari$^3$,
             Annalisa Bonafede$^1$,
             Marcus Br\"{u}ggen$^1$,
             Claudio Gheller$^4$,
             Robert Braun$^5$,
             Shea Brown$^6$
\\
	$^1$Hamburg University (Hamburg Observatory), Gojenbergsweg 112, 21029, Germany;
        $^2$INAF, Istitutio di Radioastronomia, via Gobetti 101, 40131, Bologna, Italy;
        $^3$Observatoire de la C\^{o}te d'Azur, Boulevard de l'Observatoire, B.P. 4229,  06304 Nice, France;
      $^4$CSCS, Via Trevano 131, CH-6900 Lugano, Switzerland;
      $^5$SKA Organisation, Jodrell Bank Observatory, Macclesfield, Cheshire SK11 9DL, UK;
      $^6$Department of Physics and Astronomy, University of Iowa, 203 Van Allen Hall, Iowa City, USA
     \\
    E-mail:\email{franco.vazza@hs.uni-hamburg.de}
 }

\abstract{{\bf Abstract}  The detection of the diffuse gas component of the cosmic web remains a formidable challenge. In this work we study synchrotron emission from the cosmic web with simulated SKA1 observations, which can represent an fundamental probe of the warm-hot intergalactic medium. We investigate 
radio emission originated by relativistic electrons accelerated by shocks surrounding cosmic filaments, assuming diffusive shock acceleration and as a function of the  (unknown) large-scale magnetic fields.  The detection of the brightest parts of large ($>10 \rm Mpc$) filaments of the cosmic web should be within reach of the SKA1-LOW, if the magnetic field is at the level of a $\sim 10$ percent equipartition with the thermal gas, corresponding to  $\sim 0.1 \mu G$ for the most massive filaments  in simulations.
In the course of a 2-years survey with SKA1-LOW, this will enable a first detection of the ``tip of the iceberg'' of the radio cosmic web, and allow for the use of the SKA as a powerful tool to study the origin of cosmic magnetism in large-scale structures. On the other hand, the SKA1-MID and SKA1-SUR seem less suited for this science case at low redshift ($z \leq 0.4$), owing to the missing short baselines and the consequent lack of signal from the large-scale brightness fluctuations associated with the filaments. In this case only very long exposures ($\sim 1000$ hr) may enable the detection of  $\sim 1-2$ filament for field of view in the SKA1-SUR PAF Band1.}

\FullConference{
Advancing Astrophysics with the Square Kilometre Array\\
June 8-13, 2014\\
Giardini Naxos, Italy}

\newcommand{\enzo}{\it{\small ENZO}}

\begin{document}

\section{Introduction}
The Large-Scale Structure in the Universe comprises a complex
network of filaments connecting virialized structures and
separated by voids. {\it Mildly nonlinear} structures such as sheets and filaments should contain a wealth of information on the emergence of 
cosmic structures  as  individually recognizable objects (e.g. galaxy groups and galaxy clusters).
The observational detection of the {\it warm-hot intergalactic medium} associated with filaments is still challenging, and only a few works claimed a detection in X-rays \citep[e.g.][]{2008SSRv..134...25R}, or more recently with the Sunyaev Zeldovich effect \citep[][]{2013A&A...550A.134P}.  
In  the radio waveband, a few cases of radio emission at the possible
crossroad of cosmic filaments have been reported \citep[][]{2002NewA....7..249B,2007ApJ...659..267K,2008A&A...481L..91P,2010A&A...511L...5G,2013ApJ...779..189F,gg14}.
However, such sources are likely to trace merger shocks in a filamentary environment, rather than {\it stationary accretion shocks} surrounding filaments. Numerical simulations suggest that filaments are surrounded by nearly stationary accretion shocks ($M\geq 10$), where the baryonic gas is shock-heated for the first time  \citep[][]{ry03,va11comparison}, and the detection of these shocks would confirm a critical piece of the warm-hot intergalactic medium (WHIM) paradigm \citep[e.g.][]{2001ApJ...552..473D}.
The expectations on the magnetic fields are more uncertain and lie in the range of $\sim 1 \rm nG-0.1 \mu G$, depending on numerical resolution of the cosmological simulation and as well as on the assumed seeding scenario \citep[e.g.][]{do08,ry08,donn09,va14mhd}. Provided that diffusive shock acceleration (see below) and large enough magnetic fields are present in filaments, the SKA could make the first detection of the cosmic web in the radio window \citep[][]{2004ApJ...617..281K,2011JApA...32..577B,2012MNRAS.423.2325A}.

\begin{figure*}
\begin{center}
\includegraphics[width=0.245\textwidth,height=0.36\textwidth]{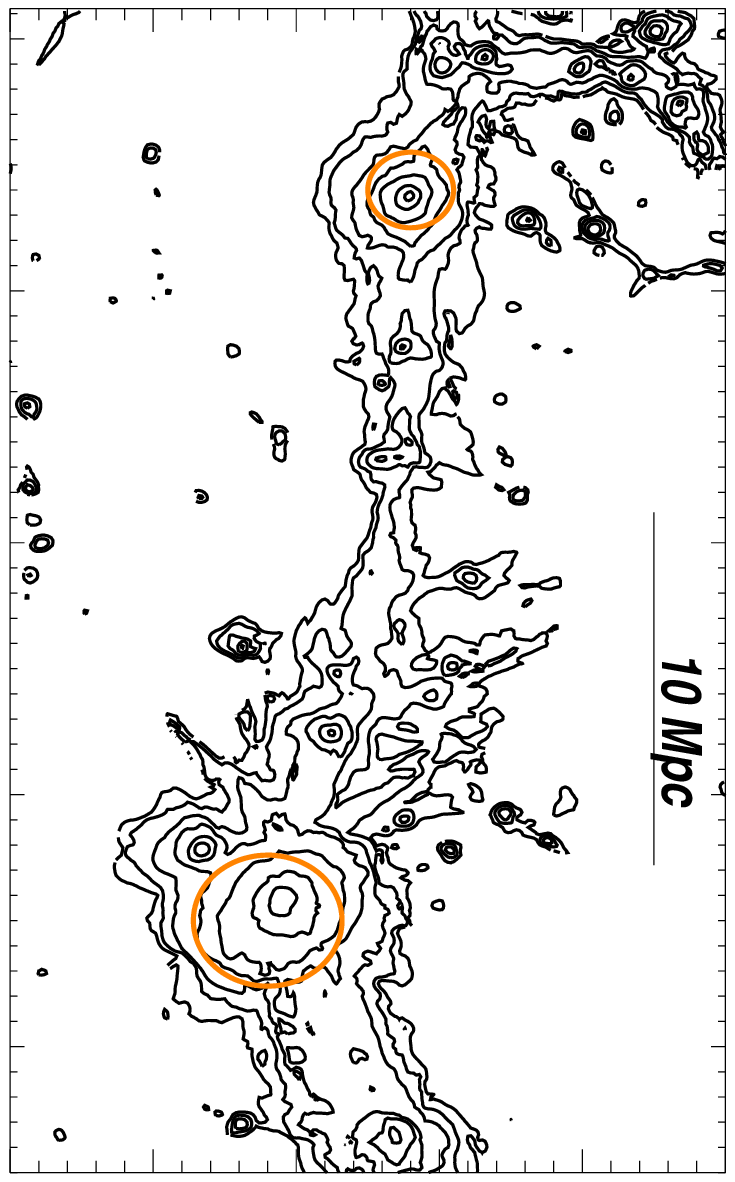}
\includegraphics[width=0.245\textwidth,height=0.36\textwidth]{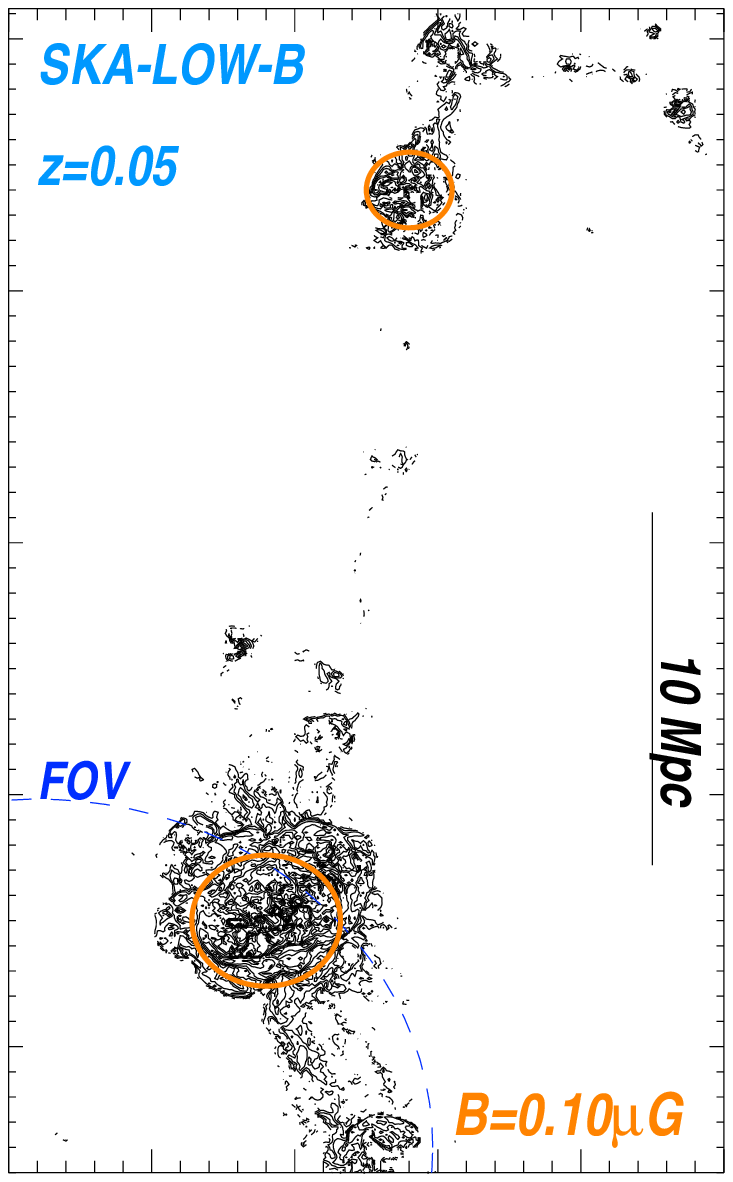}
\includegraphics[width=0.245\textwidth,height=0.36\textwidth]{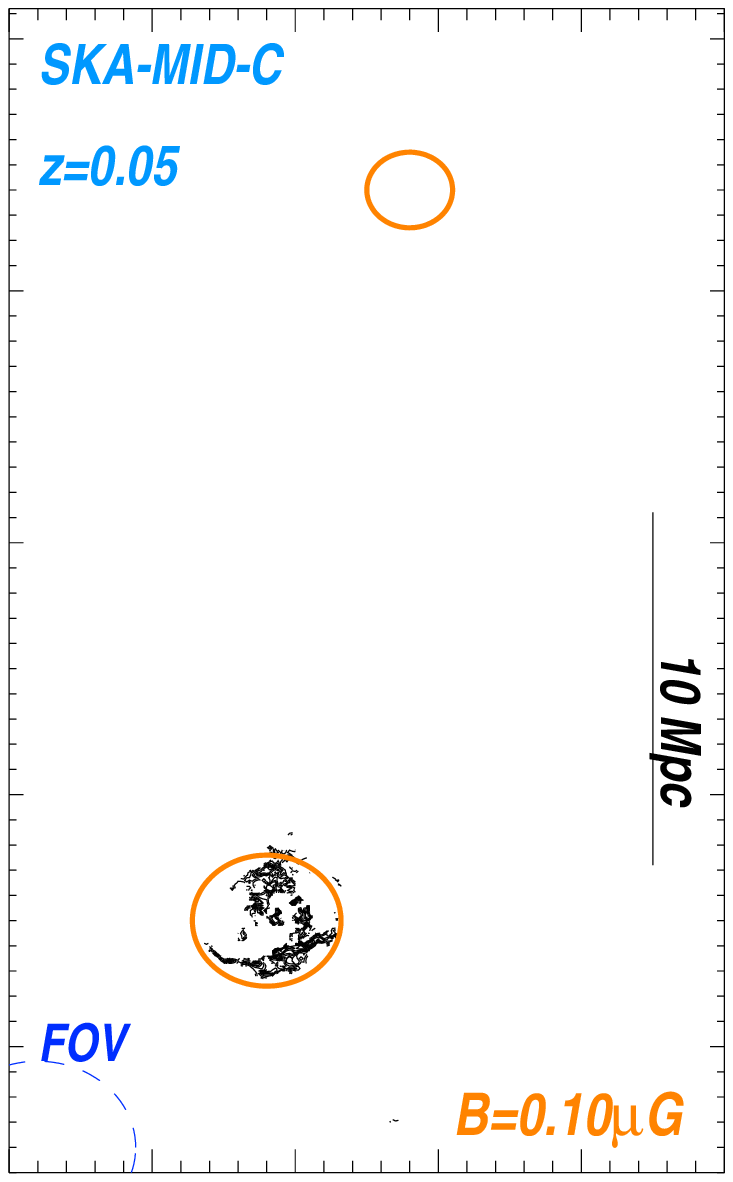}
\includegraphics[width=0.245\textwidth,height=0.36\textwidth]{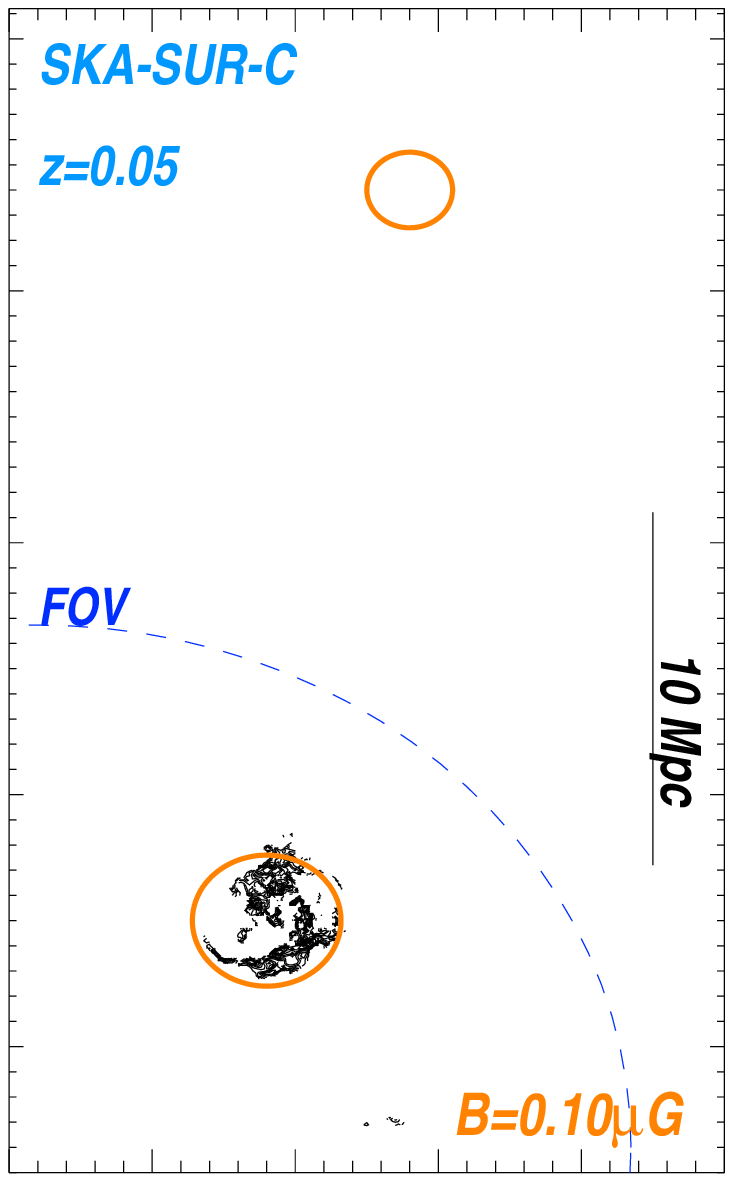}
\includegraphics[width=0.245\textwidth,height=0.36\textwidth]{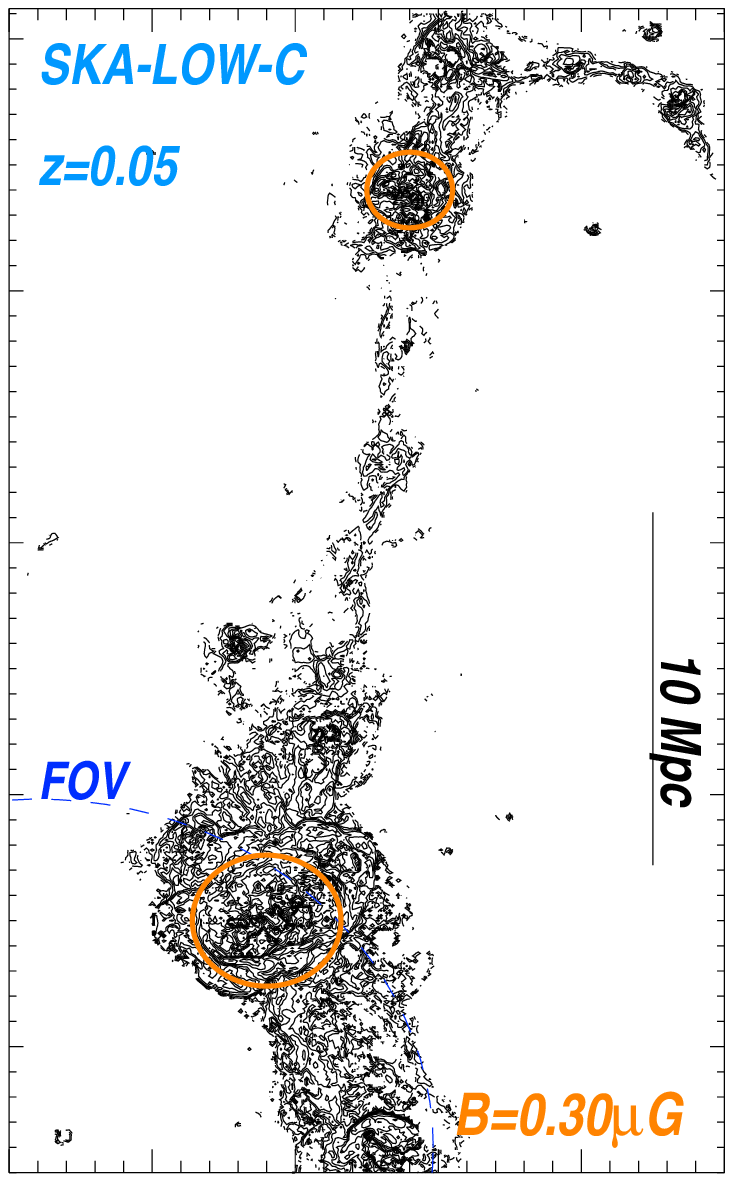}
\includegraphics[width=0.245\textwidth,height=0.36\textwidth]{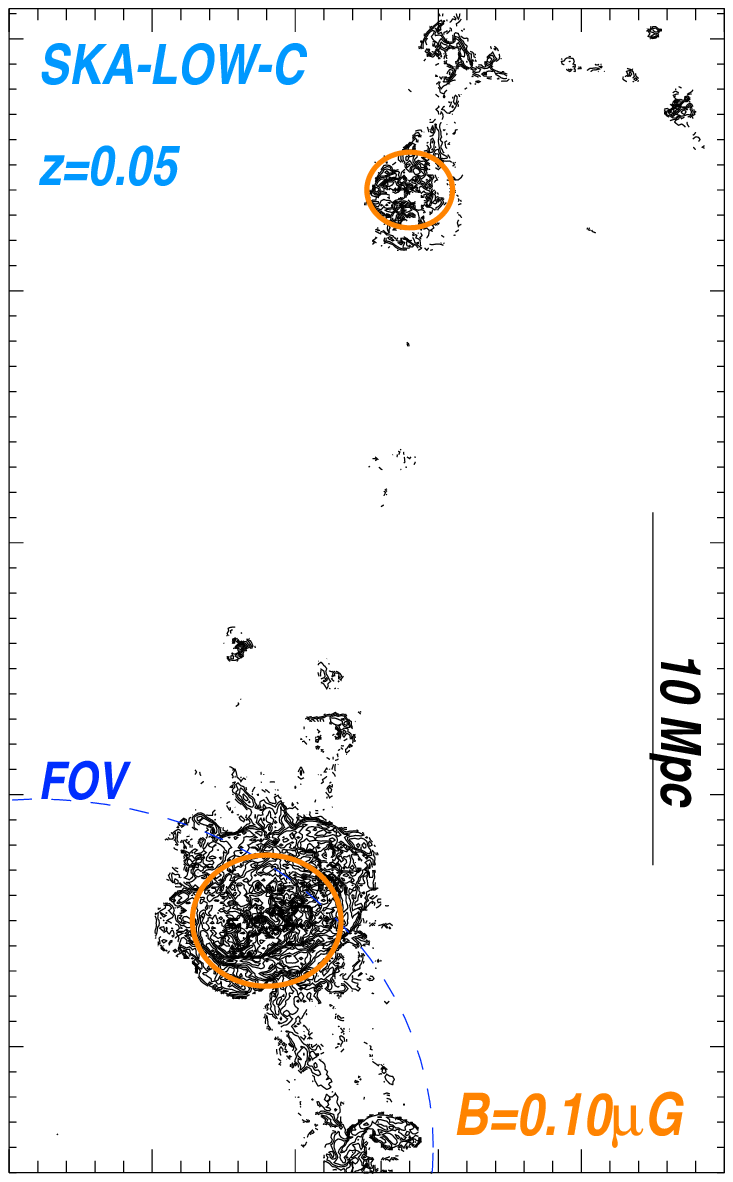}
\includegraphics[width=0.245\textwidth,height=0.36\textwidth]{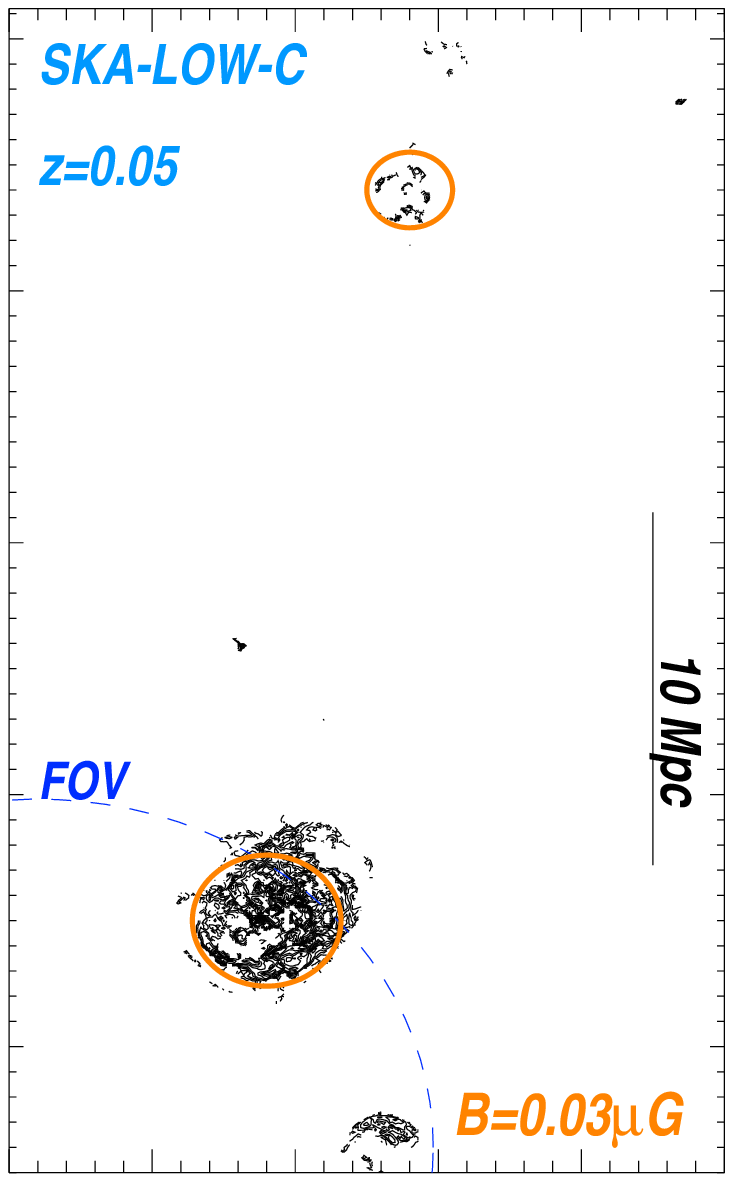}
\includegraphics[width=0.245\textwidth,height=0.36\textwidth]{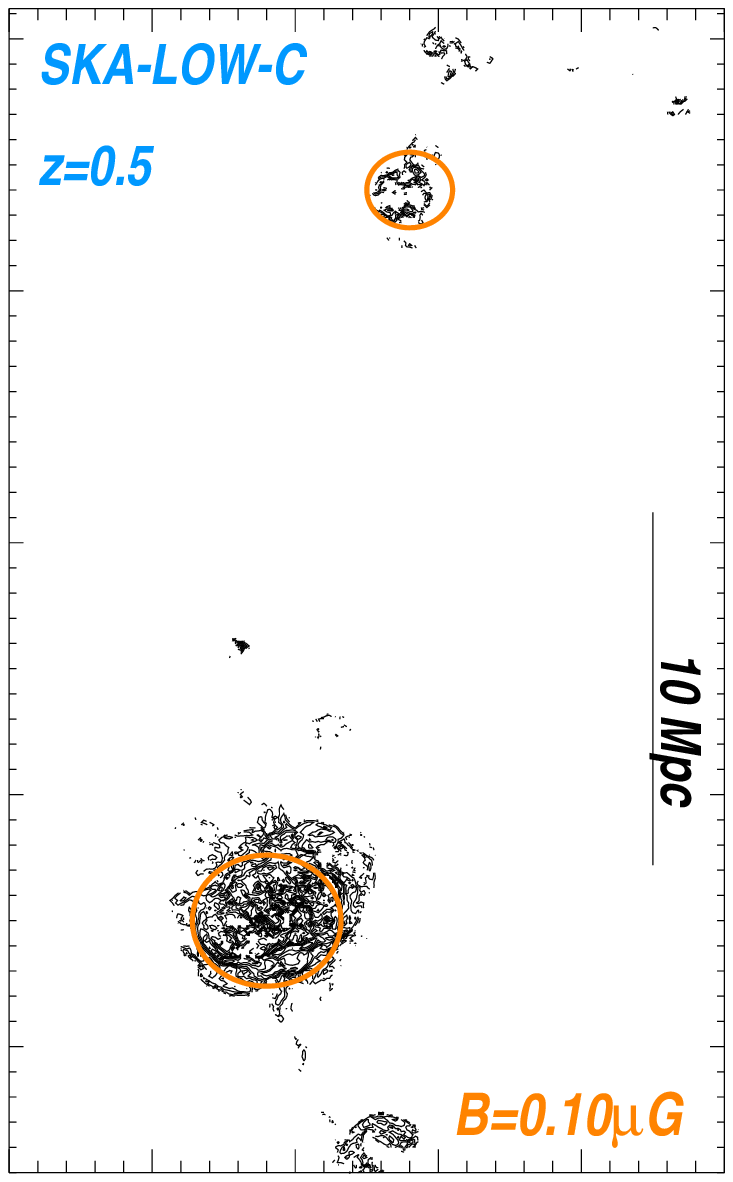}

\caption{First panel: project gas pressure map for a massive simulated filaments . Other panels: mock radio observations  with SKA1-LOW, SKA1-MID and SKA1-SUR. The top panels shows the result for  $z=0.05$ and $B=0.1 \mu G$. The lower panels are taken for the SKA1-LOW-C configuration (2 years survey at $300$ MHz and shows the result for different magnetic field strengths ($B=0.3$, $0.1$ and $0.03 \mu G$) or redshift ($z=0.05$ or $z=0.5$). The additional blue circles represent the field of view of each array (not shown if larger than the image), while the red circles show the virial radius of galaxy clusters in the image. The contours ($log_{\rm 10}[\rm Jy/arcsec^2]$) start from 3 times the nominal surface brightness (confusion) threshold for each array.}
\label{fig:fila3}
\end{center}
\end{figure*}
\section{Simulated emission}

In \citet{scienzo14} we produced large cosmological simulations with grid code ENZO \citep[][]{enzo14}, designed to study the dynamic properties of cosmic rays (CR) accelerated by cosmological shock waves. Here we mostly focus on a very large filament in our simulated volume, about $\sim 20 \rm Mpc$ in length and connecting two galaxy clusters (Fig.1). Given the size of this object, it gives us a first  gross estimate of the chances of detecting filaments with the SKA1.
Our forecasts of radio emission include two mechanisms: a) primary electron emission from relativistic electrons accelerated at shocks, assuming efficient diffusive shock acceleration (DSA, same model as in \citealt[][]{hb07}, Eq. 32); b) secondary electron emission from the relativistic electrons continuously injected by hadronic collisions in the filament volume \citep[][]{de00}, which are a by-product of our simulations \citep[][]{scienzo14}. In general, the contribution from secondary electrons is everywhere negligible, i.e. only a few percent of the primary emission.\\
The two main unknowns here are the electron acceleration efficiency at shocks ($\xi_e$) and the magnetic field within the filament ($B$); a  discussion on
these assumptions is given in Sec.\ref{sec:discussion}. 
The emitted radio power per frequency, $\nu$, scales as $I_{\nu} \propto S \cdot n \cdot \xi_e \cdot \nu^{-\delta/2} \cdot T^{3/2} \frac{B^{1+\delta/2}}{(B_{\rm CMB}^2+B^2)}$ (where $n$ and $T$ are the post-shock density/temperature, $S$ is the shock surface, and $\delta$ is the slope of the CR-electrons energy spectrum). We can expect cosmic filaments to be a sensitive probe of cosmic magnetic fields, since for shocks in filaments (where $\delta \approx 2$ and $B \ll B_{\rm CMB}$) the previous formula reduces  to  $I_{\nu} \propto \xi_e \cdot B^{2}$.
In the following we assume $\xi_e=10^{-4}$ and {\it uniform} values of $B=0.01-0.3 \mu G$ within the filament volume, and compute the detectable emission by SKA1-LOW (at two frequencies), SKA1-MID and SKA1-SUR. For each array we considered different possible strategies, including long 1000 hr exposures, 2-years surveys and the adoption point source subtraction at high resolution ($\sim 1"$), followed by tapering on point-source subtracted data (at $\sim 10"$) to have a reduced resolution (Table 1), since spatial-scale smoothing of the point-source subtracted data could significantly increase the signal-to-noise of the diffuse emission, though the power of identifying individual shock-structures would be lost \citep[][]{2011JApA...32..577B}.
For the SKA1-LOW observations we considered two possible central frequencies ($110$ and $300$ MHz) and computed the confusion noise assuming a spectral index of $-0.7$, at full resolution. In the case of SKA1-MID and SKA1-SUR, we produced forecasts for a 1000 hr exposure or a 2 years survey.
In all images we started from the full emission map obtained from the cosmological simulation, and then
selected the specific  frequency, field of view (FOV), resolution and sensitivity expected for each observation strategy, based on the SKA1 Baseline Design of March 2013 and on the SKA Phase 1 Performance Memo. In addition, we FFT-transformed all images and removed spatial frequencies smaller than the minimum baseline of each configuration, in the Fourier domain, so that the structures with a too large angular scale compared to the chosen array are removed. In all cases, we implicitly assume a perfect removal of the Milky Way foreground.

\begin{table*}
\begin{center}
\caption{Parameters for the investigated SKA1 arrays and imaging strategies, based on the SKA1 Baseline Design and on the SKA Phase 1 Performance Memo. For the tapering cases, we considered  a point source subtraction at $\sim 1"$, followed by tapering on point-source subtracted data at $\sim 10"$. The sensitivities are expressed both as a function of the beam (1) and of the sky area (2); in the case of SKA1-LOW the sensitivity is set by the confusion noise, as in SKA Phase 1 Performance Memo by assuming a spectral index of $-0.7.$.  For the SKA1-SUR and SKA1-MID we have considered a $30$ percent fractional bandwidth for a 2 years all-sky survey, while for the SKA1-LOW we have considered  an all-sky survey down to confusion noise, estimated from the rms confusion brightness at 1.4 GHz (Condon, private communication), with a confusion noise assumed as above.} 
\centering
\resizebox{\textwidth}{!}{%
\begin{tabular}{|c|c|c|c|c|c|c|c|}

 \hline 
    array & configuration/strategy &  frequency & beam size & field of view & min. baseline & sensitivity 1 & sensitivity 2 \\
          &  & $[\rm MHz]$ & $[\rm arcsec]$ & $[\rm degrees^2]$ & $[\rm m]$ & $\rm [\mu Jy/beam]$ & $\rm [\mu Jy/arcsec^2]$\\
\hline
    SKA-LOW-A & full.res+conf., survey & 110 & 6 & 8 & 45 & 5.5 & 0.134 \\
    SKA-LOW-B & full.res+conf,1000 hr & 300   & 3 & 8 & 45  & 0.5 & 0.0490 \\
    SKA-LOW-C & full.res+conf., 2 years & 300 & 10 & 8 & 45  & 9 & 0.0794 \\

  SKA-MID-A &1000hr & 1400 & 0.5 & 0.49 & 15 &  0.09 & 0.317\\
  SKA-MID-B & 2 years&1400 & 0.5 & 0.49 & 15 &  5.8 & 20.475\\ 
  SKA-MID-C & 1000 hr+tapering&1400 & 10 & 0.49 & 15 & 0.14 & 0.001\\ 
  SKA-MID-D & 2 years+tapering&1400 & 10 & 0.49 & 15 & 8.8 & 0.077\\

  SKA-SUR-A &1000hr &1200 & 1.0 & 18 & 15 & 0.4 & 0.353\\
  SKA-SUR-B & 2years&1200 & 1.0 & 18 & 15 & 3.8 & 3.353\\
  SKA-SUR-C & 1000hr+tapering&1200 & 10 & 18 & 15 & 0.5 & 0.004\\
  SKA-SUR-D &2years+tapering &1200 & 10 & 18 & 15 & 4.9 & 0.043\\

 \hline 
\end{tabular}}
\end{center}
\end{table*}

\section{Results}

\begin{figure}
\vspace{90pt}
\begin{picture}(90,90)
\put(80,40){\includegraphics[width=0.17\textwidth]{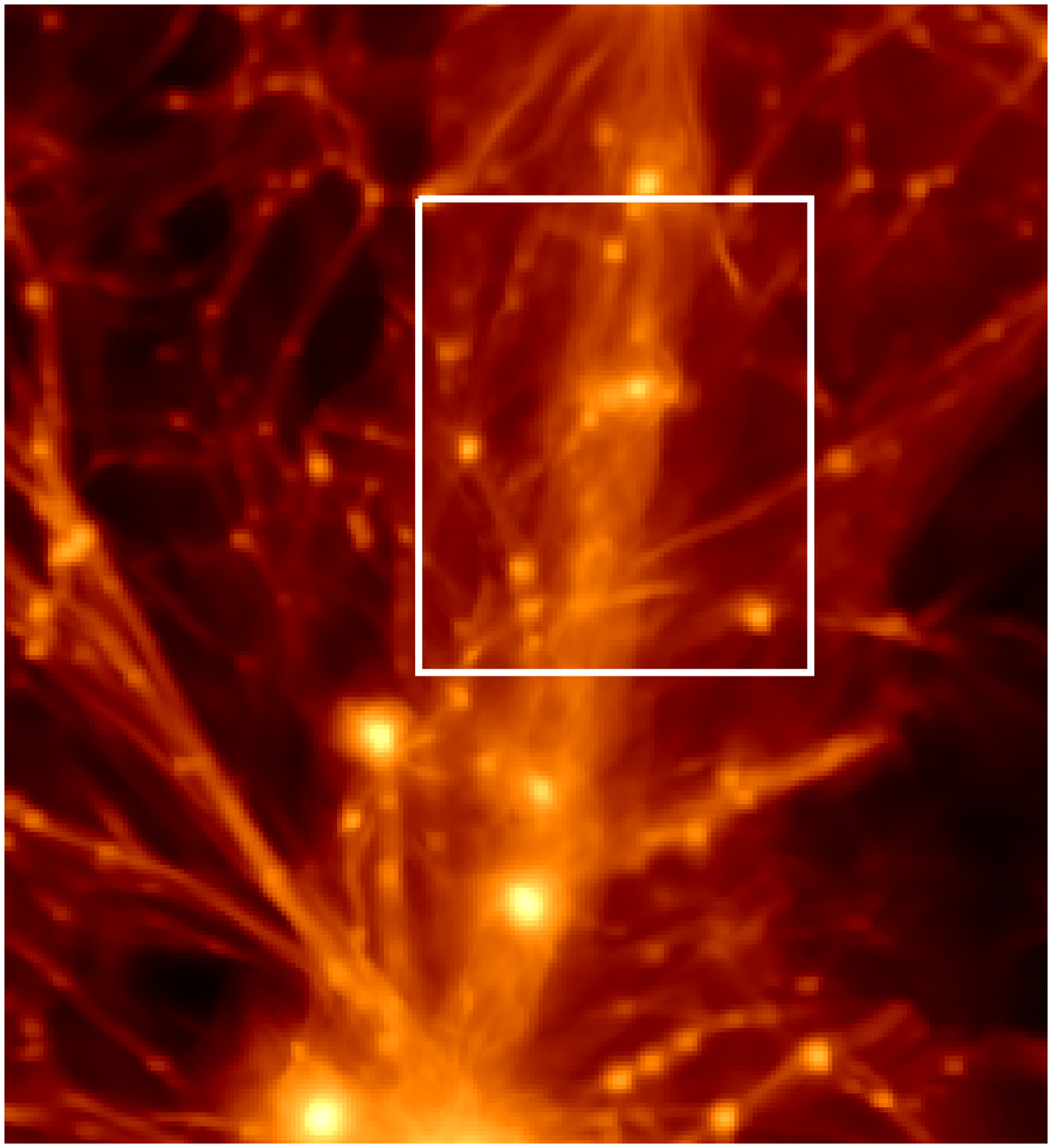}}
\includegraphics[width=0.45\textwidth,angle=90,height=0.4\textwidth]{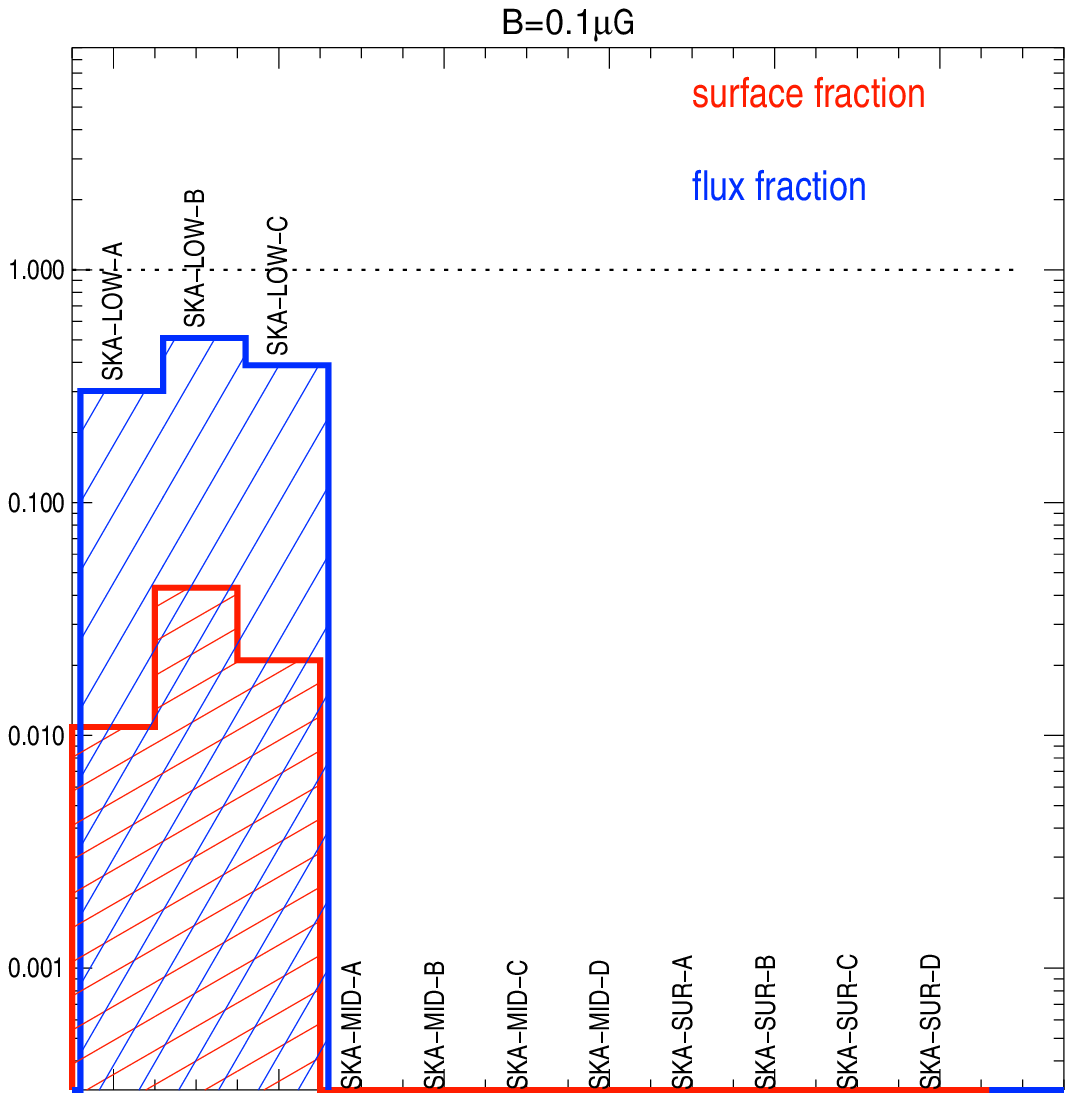}
\includegraphics[width=0.45\textwidth,angle=90,height=0.4\textwidth]{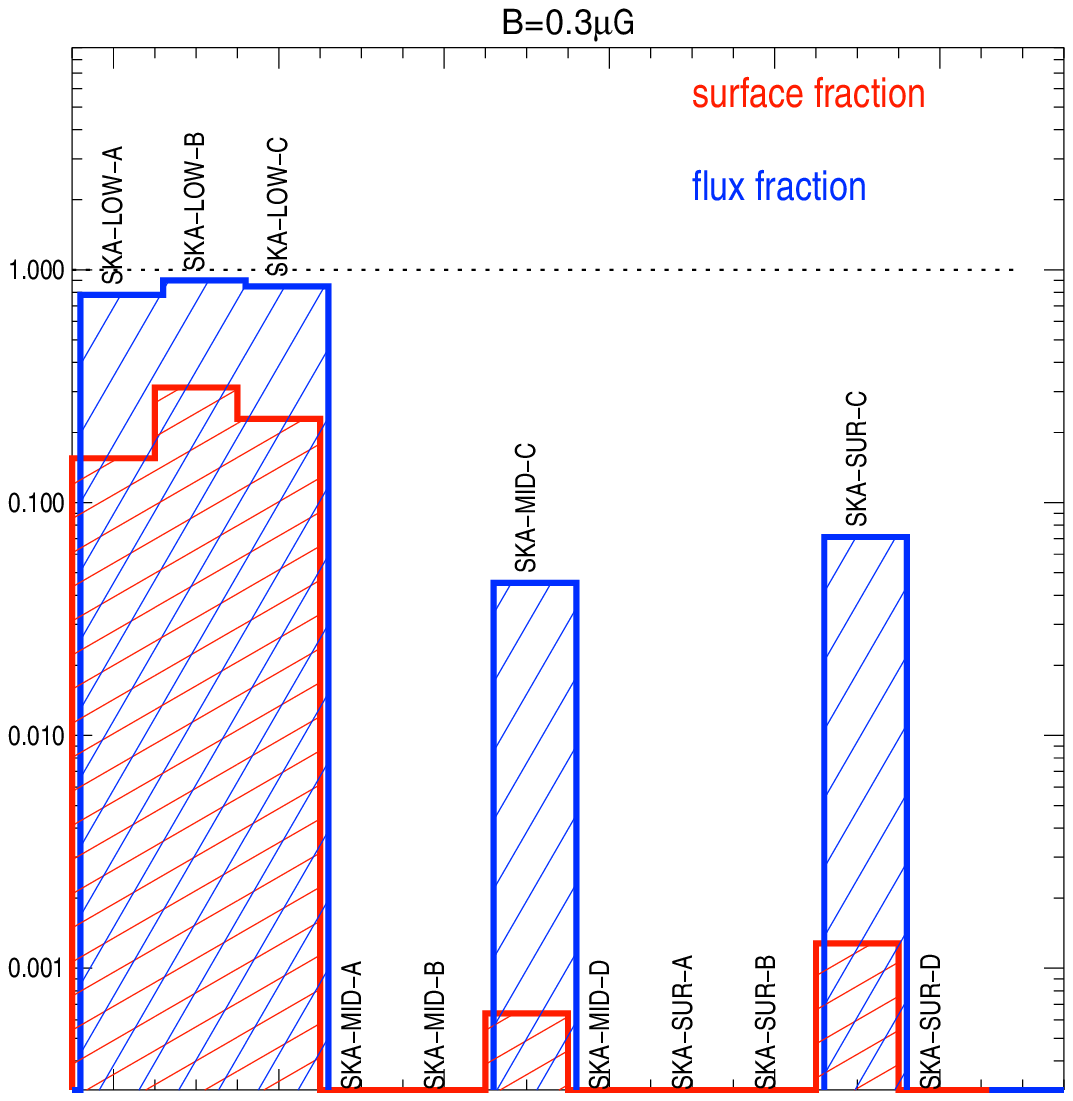}
\end{picture}
\caption{Fraction of the detectable area (red) and flux (blue) from the simulated filament, for $B=0.1 \mu G$ (left) and $B=0.3 \mu G$ (right). For a ``detection'' we considered a signal $\geq 3 \sigma$, considering the sensitivites of Table 2. The inset shows the projected gas density of our filament and the selected white area used to compute the fraction of detectable radio flux.}
\label{fig:fila6}
\end{figure}

Figure 1 shows a representative set of mock observations with various array configurations of the SKA Phase 1. The top panels, show the performances of the arrays in ``deep exposure'' mode (1000 hr), for a 
redshift $z=0.05$ and assuming $B=0.1 ~\mu G$ \footnote{We remark that in this test we are only concerned with the possible emission from the WHIM from the filament, and that this emission model is likely overestimating the radio emission within cluster (e.g. radio relic emission), because of the assumed {\it constant} electron acceleration efficiency and magnetic fields.}. The SKA1-LOW-B array should detect at least a significant portion of the filament at $3 \sigma$, and trace some of the brightest emission knots
connecting the two clusters in the field, as well as the big filamentary accretion southwards of the most
massive cluster, reaching out to $\sim 5 ~\rm Mpc$ from the virial cluster volume.
Instead, with the SKA1-MID and SKA1-SUR arrays the detection of the filament will be impossible even with a 1000 hr pointed observation, because of the smaller sensitivity and the lack of short baselines (at least the central frequencies of $1.4$ and $1.2$ GHz assumed here). The situation gets even worse by considering the shallower sensitivity reached in the 2-year survey.
However, even with the SKA1-LOW the extended emission from the filament connecting the two clusters would be barely detectable in its entire extent only for a higher magnetic field, e.g. $B=0.3 \mu G$. 
A non-detection of such structures at low redshift with the SKA1-LOW, would place robust upper-limits on the value of the intergalactic magnetic fields, of the order of $\leq 0.03 \mu G$, giving us important cluse on the efficiency of magnetic field amplifications in the WHIM.
At larger redshifts ($z \geq 0.1$), the cosmological dimming makes the detection impossible, and also the removal of the contribution from single radio galaxies becomes challenging. \\
In order to better quantify the detectability of the simulated filament in the different arrays and using different possible imaging strategies, we selected a smaller region inside the main filament (see the map in Fig.\ref{fig:fila6}), well outside the surrounding clusters outskirts, in order not to be contaminated by cluster accretion shocks. 
We computed the total area of the source region above the $3 \sigma$ surface brightness level, as well as the ratio between the detectable flux and the total one from the same region, as shown in Fig.\ref{fig:fila6}, for $B=0.3 \mu G$ or $B=0.1 \mu G$. \\
If $B=0.1 \mu G$ the SKA1-LOW is the only one giving any chance of detection. Using the planned array configuration for the Phase 1 of SKA1-LOW our forecast is that $\sim 30-60$ percent of the total flux from the filament would be detectable, if most of this emission came from the brightest and densest gas knots within the filament rather than
from the diffuse WHIM. With the deep exposure of SKA1-LOW at $300$ MHz (case B) the percentage of detectable flux approaches to $\sim 60$ percent, while the detectable surface goes to $\sim 5$ percent of the total.\\
Based on the same computations, we can esimate that an increase in sensitivity of a factor $\times 3-5$ will
enable a clear detection of the full emission from such filament, up to several $\rm Mpc$ from the cluster centre, as well as the detection of the brightest emission spots from the filament at $z=0.5$. 
This will be feasible in the Phase 2 of the SKA1-LOW, since the planned $\times 20$ increase of the maximum angular resolution (down to $\sim 0.15``$) in the $50-350 ~\rm MHz$ range will lead to a great decrease of the confusion noise. \\
On the other hand, in the case of SKA-MID and SKA-SUR even in Phase 2 the lack of short baseline would make it impossible to detect the bulk of synchrotron emission from the largest scales in filaments at the central frequencies we studied here.

\section{Discussion}
\label{sec:discussion}
The forecast of radio emission from the diluted gas of cosmic filaments strongly depends on the combined 
value of $\xi_e \cdot B^2$, both of which uncertain.
\begin{itemize}
 \item {\it The electron acceleration efficiency}. The qualitative similarity to strong shocks in supernova remnants suggests that for the acceleration efficiency of electrons at strong shock $\xi_e \approx 10^{-4}$ can be assumed \citep[e.g.][]{ed11}. However, the extrapolation of the DSA to the plasma condition of filaments is uncertain, since a too weak degree magnetisation would inhibit the formation of the magnetic field irregularities responsible for efficient
particle scattering. Recent hybrid simulations of collisionless shocks suggest that this mechanism
can still operate in presence of very low magnetisation, owing to the fast development of micro-instabilities at parallel shocks \citep[][]{ca14}.  In general, {\it any} detection of radio emission associated to filament shocks will inform us that DSA (or some modified version of the theory) can operate down to this low density regime ($n \leq 10^{-5} \rm cm^{-3}$), carrying an important theoretical information. \\
\item {\it The magnetic field}. The expectations of MHD cosmological simulations lie in the range of $\sim 0.001-0.1 \mu G$ in filaments, depending on the numerical resolution and on the seeding scenarios. In the case of the giant filament analysed in this work, a value of $B=0.1 \mu G$ correspond to a magnetic energy density at the level of $\sim 10$ percent of the gas energy, which seems very hardly achievable by small-scale amplification of primordial seed fields alone \citep[][]{br05, ry08, va14mhd}. However, many additional astrophysical seeding scenarios of magnetic fields might add to the cosmological seed field \citep[e.g.][]{donn09}. Any  detection of coherent emission on $\geq 1-10 ~\rm Mpc$ scales in filaments will rule out the scenario in which the magnetisation  of filaments is driven by low-redshift seeding of magnetic fields by galactic activity, owing to the difficulty of covering such large scales with outflows or diffusion. At the same time, the detection of synchrotron from filaments at high redshift will give us important clues on the efficiency of turbulent amplification of magnetic fields over cosmological epochs.
\end{itemize}

Some insight can also be gained from the fact that no spectacular large-scale filament has been detected so far by LOFAR Cycle 0 observations.
The present non-detection can be used to adjust our forecasts for SKA1. Following this conservative approach, we simulated the LOFAR LBA and HBA observations considering $\approx 8 ~\mu J/\rm arcsec^2$ ($\approx 1.7 ~\mu J/\rm arcsec^2$), as achieved in the current LOFAR-LBA (HBA) observations of nearby clusters, and a $23 \times 16 ~\rm arcsec^2$ ($6 \times 6 \rm arcsec^2$) beam {\footnote{ https://www.astron.nl/radio-observatory/astronomers/array-configurations/.}}.
In order {\it not} to be detected with present LOFAR Cycle 0 observations, $B \leq 0.1 ~\mu G$, and we therefore assume this conservative value for the maximum magnetic field in giant filaments, provided that DSA
is at work in these environments. As we have seen, even in this case the SKA1-LOW can still detect a significant portion of the filament emission for $z \leq 0.1$.\\

\begin{figure}
 \includegraphics[height=0.49\textwidth,angle=90]{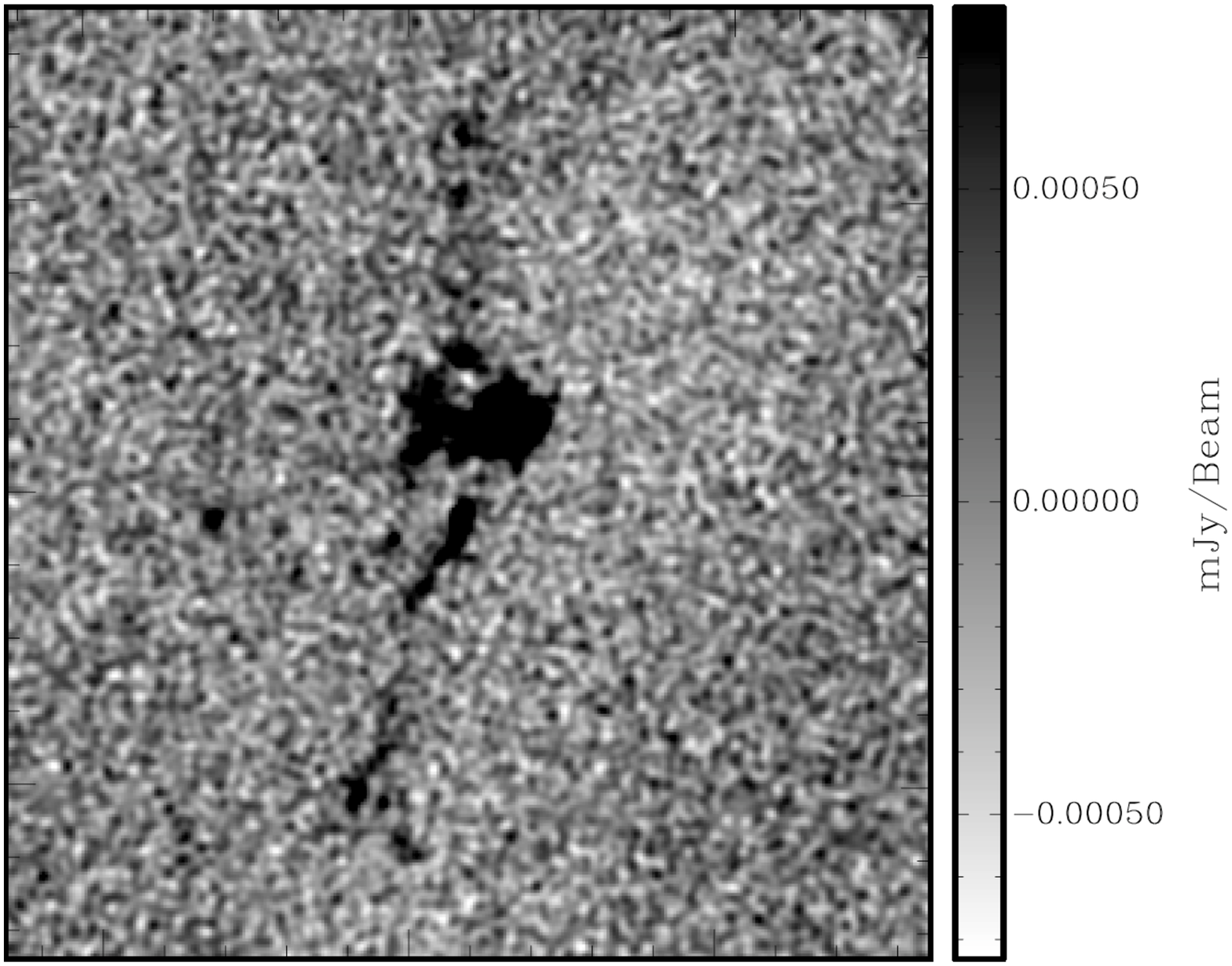}
 \includegraphics[height=0.49\textwidth,angle=90]{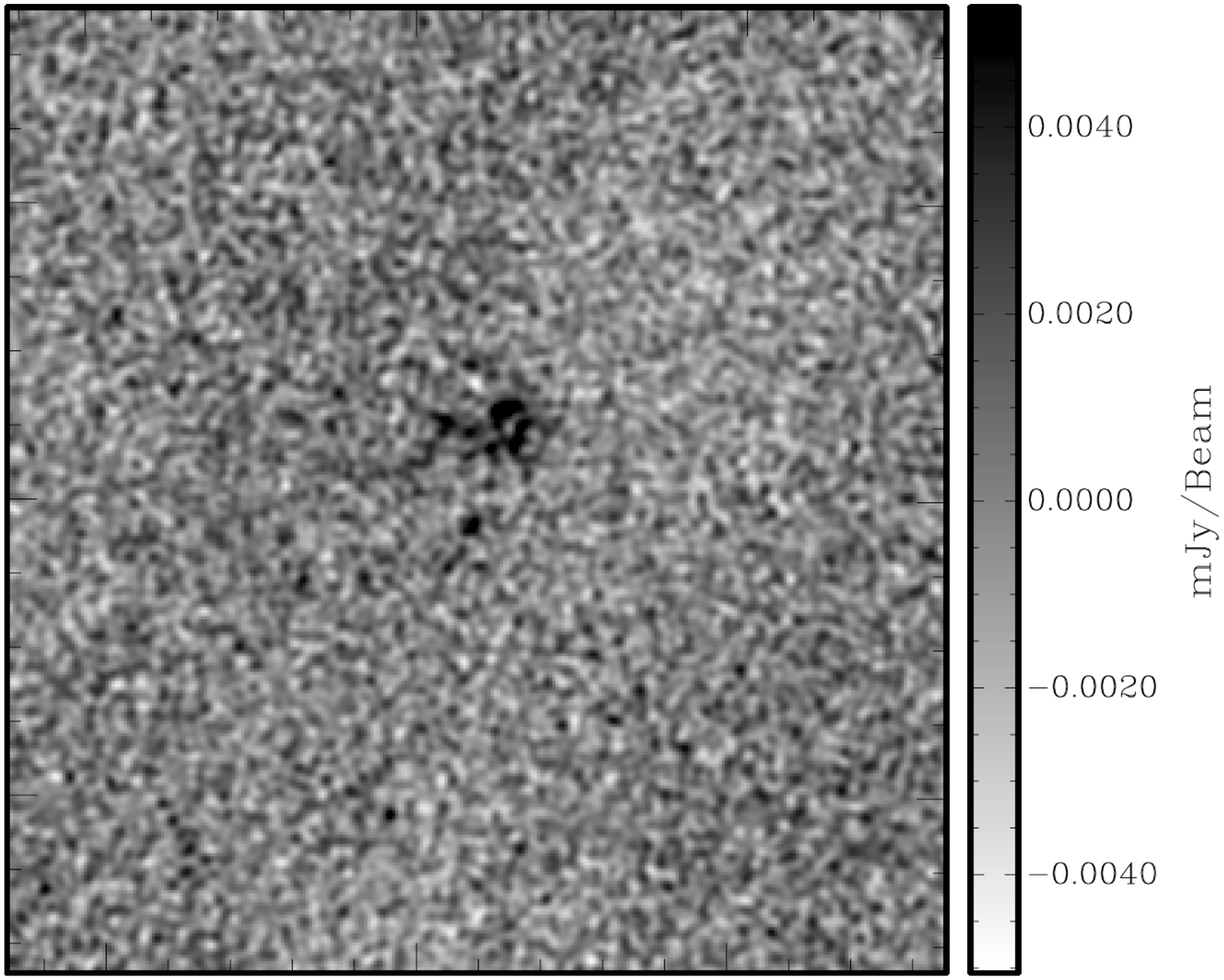}
\caption{Mock observations ofa $\approx 5$ degree area within the filament. (as in the white selection of Fig.2, assuming a distance of $z=0.02$), using the SKA simulator: 2000 hr pointing (left) and 60 hours pointing (right) at 140 MHz with the SKA1-LOW (see text for more details).}
\end{figure}


\begin{figure}
\begin{center}
\includegraphics[width=0.99\textwidth]{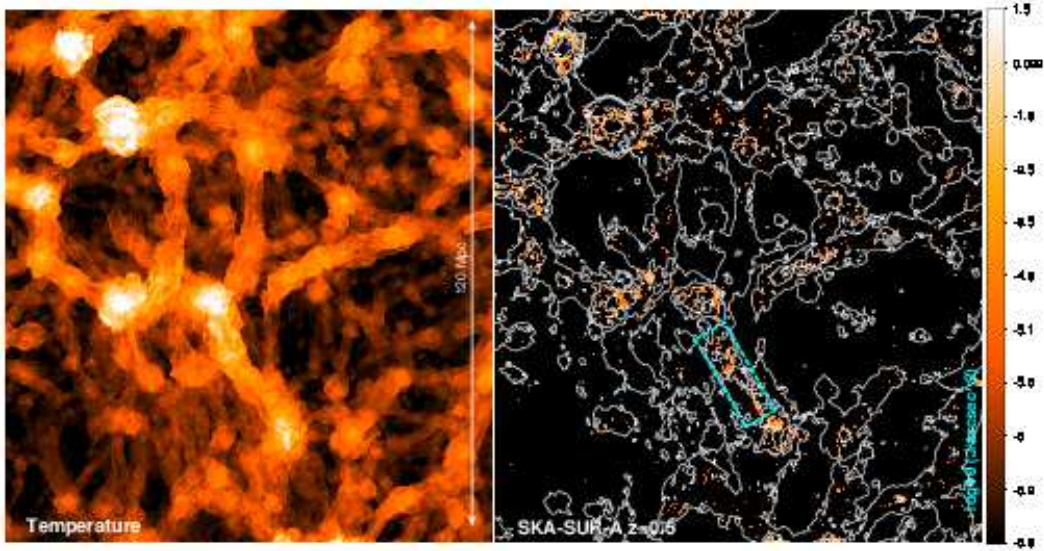}
\caption{Left panel: projected gas temperature of a large cosmological volume at $z=0.5$. Right panel: detectable emission by SKA1-SUR-A, using a 1000 hr exposure. The emission coming from inside the virial radius of all simulated clusters has been blanked out. The rectangular selection highlight a giant filament whose detection appears within reach.}
\label{fig:fila7}
\end{center}
\end{figure}

\section{Conclusions}

The detection of large cosmic filaments in radio at low redshift seems to be within reach by the SKA1-LOW, provided that the massive filaments are magnetised at the level of $B \sim 0.1 ~\mu G$ ($\sim 1-10$ percent of equipartition with the thermal gas energy) and that DSA is at
work in the tenous WHIM environment. The SKA1-LOW is always favoured because of its better sensitivity and 
sampling of short baselines compared to SKA1-MID and SKA1-SUR.
In particular, the SKA1-LOW should be able to detect a significant portion of filament emission at low redshifts ($z \leq 0.1$), while for a better imaging of their full extent or for their detection up to $z=0.5$ (or for lower magnetic fields) its sensitivity should be improved by a factor $\sim 3-5$. This can be achieved during the Phase 2 of SKA-LOW, thanks to the planned increased in the maximum 
angular resolution, which will enable to beat the confusion noise. On the other hand, our tests show that for most of cosmic filaments in the nearby ($z \leq 0.1 $) Universe a detection with SKA1-MID or SKA1-SUR is made impossible by the lack of short baselines, for which even Phase 2 will not be beneficial. 
  
This work represents just a first step towards the realistic modelling of radio imaging of the
cosmic web with the SKA1 and the SKA2. Our schematic procedure of removing the missing
baselines in the the Fourier plane is a very crude estimate of the actual UV coverage, and the
effect of noise, confusion, finite bandwith etc. are not yet properly accounted for. In order to fill
this gap, we produced a first set of simulated observations of our models with the most updated version of
the simulated SKA performance \footnote{http://astronomers.skatelescope.org/wp-content/uploads/2014/06/SKA1\_Science}. In Figure 3 we present single pointing images at $140$ MHz of a $\approx 5$ degree
field of view centred on the faint filament region at $z = 0.02$ selected in Fig.\ref{fig:fila6}, for the $B = 0.1 ~\rm \mu G$ case. For the left image we used a $2000$ hours of
integration to achieve
$\approx  220 ~\rm nJy/beam$ with $30$ percent fractional bandwidth, while for the right
image we simulated a
$60$ hour integration, achieving $\approx 1.3 ~ \rm \mu Jy/beam$ (in both cases using a $300$
arcsec beam).
The trends we recover with the use of the sophisticated tool provided by
the SKA simulator overall agree with our previous estimates: while in a shallow survey mode only the
brightest density peaks inside the filament can be marginally detected, the deep integration is
able to pinpoint also the faint elongated structure of the filament, thereby suggesting its large-scale
orientation and the presence of accretion shocks along several $\rm Mpc$.

Future deep exposures will also enhance the statistical chance of detecting even more massive filaments in large areas of the sky, not only with the SKA1-LOW if the targets are distant enough to have a sufficient baseline sampling even at higher frequencies. Figure \ref{fig:fila7} shows a test where we used a large simulated volume of $(150^3 \rm Mpc^3 )$ (using $1024^3$ cells) and simulated
the detectable emission using a long 1000 hr exposure with SKA1-SUR (SKA1-SUR-A in Tab.1) and assuming a redshift of $z=0.5$.
In this case, we attached a magnetic field to the simulation by assuming an average $5$ percent equipartition with the thermal energy within each cell. The field of view in the image approximately corresponds to 3 fields of view of the SKA1 survey Band1 ($\approx 18$ square degrees).
Most of the detectable radio emission by the SKA1-SUR will originate from the cluster outskirts, yet in a few cases also the ''tip of the iceberg`` of radio emission coming from very big filaments should be detectable, like in the case of the hot filament within the rectangle in Fig.\ref{fig:fila7}. Our tests suggest that the redshift range $0.4 \leq z \leq 0.5$ is the best to have some chance of detection with the  SKA1-SUR PAF Band1, while at $z<0.4$ and $z>0.5$ the emission will be filtered-out and suppressed by cosmological dimming, respectively. Based on a crude extrapolation of our data, we expect a detection of $\sim 1-2$ filaments of such giant filaments for every FOVs of SKA1-SUR, using $\sim 1000$ hr of exposure. Deep exposures in {\it polarisation} should maximise the chance of detecting these rare structures, since they are expected to be highly polarised owing to their strong Mach numbers. Polarised observations are not dynamical-range limited, and warrant that the thermal noise
will be reached independently of the presence of strong source within the field of view, thereby increasing the chances of imaging such low-surface brightness structures in the cosmic web.

\section*{Acknowledgements}

The computations described in this work were performed using the {\enzo} code (http://enzo-project.org).
We acknowledge PRACE for awarding us access to CURIE-Genci based in France at Bruyeres-le-Chatel. The support of the TGC Hotline from the Centre CEA-DAM Ile de France to the technical work is gratefully acknowledged. We also acknowledge CSCS-ETH for the use of the Cray XC30 Piz Daint.
F.V., M.B. and A.B. acknowledge support from the grant FOR1254 from the Deutsche Forschungsgemeinschaft. We thank the ``Tiger Team'' of SKA-LOW and J. Condon for useful scientific discussion, and we gratefully acknowledge the assistance by the SKA Office to produce simulated SKA observations.

\bibliographystyle{apj}
\bibliography{franco}

\begin{thebibliography}{}
\expandafter\ifx\csname natexlab\endcsname\relax\def\natexlab#1{#1}\fi

\bibitem[{{Araya-Melo} {et~al.}(2012){Araya-Melo}, {Arag{\'o}n-Calvo},
  {Br{\"u}ggen}, \& {Hoeft}}]{2012MNRAS.423.2325A}
{Araya-Melo}, P.~A., {Arag{\'o}n-Calvo}, M.~A., {Br{\"u}ggen}, M., \& {Hoeft},
  M. 2012, \mnras, 423, 2325

\bibitem[{{Bagchi} {et~al.}(2002){Bagchi}, {En{\ss}lin}, {Miniati}, {Stalin},
  {Singh}, {Raychaudhury}, \& {Humeshkar}}]{2002NewA....7..249B}
{Bagchi}, J., {En{\ss}lin}, T.~A., {Miniati}, F., {et~al.} 2002, \na, 7, 249

\bibitem[{{Brown}(2011)}]{2011JApA...32..577B}
{Brown}, S.~D. 2011, Journal of Astrophysics and Astronomy, 32, 577

\bibitem[{{Br{\"u}ggen} {et~al.}(2005){Br{\"u}ggen}, {Ruszkowski},
  {Simionescu}, {Hoeft}, \& {Dalla Vecchia}}]{br05}
{Br{\"u}ggen}, M., {Ruszkowski}, M., {Simionescu}, A., {Hoeft}, M., \& {Dalla
  Vecchia}, C. 2005, \apjl, 631, L21

\bibitem[{{Bryan} {et~al.}(2014){Bryan}, {Norman}, {O'Shea}, {Abel}, {Wise},
  {Turk}, {Reynolds}, {Collins}, {Wang}, {Skillman}, {Smith}, {Harkness},
  {Bordner}, {Kim}, {Kuhlen}, {Xu}, {Goldbaum}, {Hummels}, {Kritsuk}, {Tasker},
  {Skory}, {Simpson}, {Hahn}, {Oishi}, {So}, {Zhao}, {Cen}, {Li}, \& {Enzo
  Collaboration}}]{enzo14}
{Bryan}, G.~L., {Norman}, M.~L., {O'Shea}, B.~W., {et~al.} 2014, \apjs, 211, 19

\bibitem[{{Caprioli} \& {Spitkovsky}(2014)}]{ca14}
{Caprioli}, D., \& {Spitkovsky}, A. 2014, \apj, 783, 91

\bibitem[{{Dav{\'e}} {et~al.}(2001){Dav{\'e}}, {Cen}, {Ostriker}, {Bryan},
  {Hernquist}, {Katz}, {Weinberg}, {Norman}, \& {O'Shea}}]{2001ApJ...552..473D}
{Dav{\'e}}, R., {Cen}, R., {Ostriker}, J.~P., {et~al.} 2001, \apj, 552, 473

\bibitem[{{Dolag} {et~al.}(2008){Dolag}, {Bykov}, \& {Diaferio}}]{do08}
{Dolag}, K., {Bykov}, A.~M., \& {Diaferio}, A. 2008, \ssr, 134, 311

\bibitem[{{Dolag} \& {En{\ss}lin}(2000)}]{de00}
{Dolag}, K., \& {En{\ss}lin}, T.~A. 2000, \aap, 362, 151

\bibitem[{{Donnert} {et~al.}(2009){Donnert}, {Dolag}, {Lesch}, \&
  {M{\"u}ller}}]{donn09}
{Donnert}, J., {Dolag}, K., {Lesch}, H., \& {M{\"u}ller}, E. 2009, \mnras, 392,
  1008

\bibitem[{{Edmon} {et~al.}(2011){Edmon}, {Kang}, {Jones}, \& {Ma}}]{ed11}
{Edmon}, P.~P., {Kang}, H., {Jones}, T.~W., \& {Ma}, R. 2011, \mnras, 414, 3521

\bibitem[{{Farnsworth} {et~al.}(2013){Farnsworth}, {Rudnick}, {Brown}, \&
  {Brunetti}}]{2013ApJ...779..189F}
{Farnsworth}, D., {Rudnick}, L., {Brown}, S., \& {Brunetti}, G. 2013, \apj,
  779, 189

\bibitem[{{Giovannini et al.}(2015)}]{gg14}
{Giovannini}, G., et~al. 2015, in proceedings of Advancing Astrophysics with the
  Square Kilometer Array,PoS(AASKA14)104

\bibitem[{{Giovannini} {et~al.}(2010){Giovannini}, {Bonafede}, {Feretti},
  {Govoni}, \& {Murgia}}]{2010A&A...511L...5G}
{Giovannini}, G., {Bonafede}, A., {Feretti}, L., {Govoni}, F., \& {Murgia}, M.
  2010, \aap, 511, L5

\bibitem[{{Hoeft} \& {Br{\"u}ggen}(2007)}]{hb07}
{Hoeft}, M., \& {Br{\"u}ggen}, M. 2007, \mnras, 375, 77

\bibitem[{{Keshet} {et~al.}(2004){Keshet}, {Waxman}, \&
  {Loeb}}]{2004ApJ...617..281K}
{Keshet}, U., {Waxman}, E., \& {Loeb}, A. 2004, \apj, 617, 281

\bibitem[{{Kronberg} {et~al.}(2007){Kronberg}, {Kothes}, {Salter}, \&
  {Perillat}}]{2007ApJ...659..267K}
{Kronberg}, P.~P., {Kothes}, R., {Salter}, C.~J., \& {Perillat}, P. 2007, \apj,
  659, 267

\bibitem[{{Pizzo} {et~al.}(2008){Pizzo}, {de Bruyn}, {Feretti}, \&
  {Govoni}}]{2008A&A...481L..91P}
{Pizzo}, R.~F., {de Bruyn}, A.~G., {Feretti}, L., \& {Govoni}, F. 2008, \aap,
  481, L91

\bibitem[{{Planck Collaboration} {et~al.}(2013){Planck Collaboration}, {Ade},
  {Aghanim}, {Arnaud}, {Ashdown}, {Atrio-Barandela}, {Aumont}, {Baccigalupi},
  {Balbi}, {Banday}, \& et~al.}]{2013A&A...550A.134P}
{Planck Collaboration}, {Ade}, P.~A.~R., {Aghanim}, N., {et~al.} 2013, \aap,
  550, A134

\bibitem[{{Richter} {et~al.}(2008){Richter}, {Paerels}, \&
  {Kaastra}}]{2008SSRv..134...25R}
{Richter}, P., {Paerels}, F.~B.~S., \& {Kaastra}, J.~S. 2008, \ssr, 134, 25

\bibitem[{{Ryu} {et~al.}(2008){Ryu}, {Kang}, {Cho}, \& {Das}}]{ry08}
{Ryu}, D., {Kang}, H., {Cho}, J., \& {Das}, S. 2008, Science, 320, 909

\bibitem[{{Ryu} {et~al.}(2003){Ryu}, {Kang}, {Hallman}, \& {Jones}}]{ry03}
{Ryu}, D., {Kang}, H., {Hallman}, E., \& {Jones}, T.~W. 2003, \apj, 593, 599

\bibitem[{{Vazza} {et~al.}(2014{\natexlab{a}}){Vazza}, {Br{\"u}ggen},
  {Gheller}, \& {Wang}}]{va14mhd}
{Vazza}, F., {Br{\"u}ggen}, M., {Gheller}, C., \& {Wang}, P.
  2014{\natexlab{a}}, \mnras, 445, 3706

\bibitem[{{Vazza} {et~al.}(2011){Vazza}, {Dolag}, {Ryu}, {Brunetti}, {Gheller},
  {Kang}, \& {Pfrommer}}]{va11comparison}
{Vazza}, F., {Dolag}, K., {Ryu}, D., {et~al.} 2011, \mnras, 418, 960

\bibitem[{{Vazza} {et~al.}(2014{\natexlab{b}}){Vazza}, {Gheller}, \&
  {Br{\"u}ggen}}]{scienzo14}
{Vazza}, F., {Gheller}, C., \& {Br{\"u}ggen}, M. 2014{\natexlab{b}}, \mnras,
  439, 2662

\end{thebibliography}

\end{document}